%% file: neurips_2026.tex
\documentclass{article}

\PassOptionsToPackage{numbers,compress}{natbib}
\usepackage[dblblindworkshop, final]{neurips_2026}

\usepackage[utf8]{inputenc}
\usepackage[T1]{fontenc}
\usepackage{hyperref}
\hypersetup{hypertexnames=false}
\usepackage{url}
\usepackage{booktabs}
\usepackage{amsmath,amssymb,amsfonts}
\usepackage{nicefrac}
\usepackage{microtype}
\usepackage{xcolor}
\usepackage{graphicx}
\usepackage{multirow}
\usepackage{float}
\usepackage{silence}
\title{
\vspace{-4pt}
Adaptive-Shot Hybrid Quantum Anomaly Detection for Tactile Internet Security: Reliability-Aware Measurement Allocation Under Resource Constraints
\vspace{-4pt}
}

\workshoptitle{SaTQuML: Secure and Trustworthy Quantum Machine Learning}

\author{%
  Mubassir Serneabat Sudipto$^{1}$\\
  {\small $^{1}$College of Engineering}\\
  {\small Iowa State University}\\
  {\small Ames, Iowa, United States}\\
  {\small \texttt{msudipto@iastate.edu}}
  \And
  Shakil Ahmed$^{2}$ \\
  {\small $^{2}$College of Computing}\\
  {\small Grand Valley State University}\\
  {\small Allendale, Michigan, United States}\\
  {\small \texttt{ahmeshak@gvsu.edu}}
  \AND
  Ashfaq Khokhar$^{3}$ \\
  {\small $^{3}$Carl R. Ice College of Engineering}\\
  {\small Kansas State University}\\
  {\small Manhattan, Kansas, United States}\\
  {\small \texttt{akhokhar@ksu.edu}}
  \And
  Samir M. Iqbal$^{4}$ \\
  {\small $^{4}$College of Computing}\\
  {\small Grand Valley State University}\\
  {\small Allendale, Michigan, United States}\\
  {\small \texttt{iqbalsa@gvsu.edu}}
}

\begin{document}

\maketitle
\vspace{-12pt}

\begin{abstract}
Tactile Internet (TI) security analytics must balance reliable thresholded decisions with constrained computational and measurement resources. We study this tension for finite-shot hybrid quantum anomaly inference and introduce the Adaptive-Shot Variational Quantum Circuit (AS-VQC) policy. This validation-calibrated policy begins each record at 128 shots and cumulatively escalates through 256, 512, and 1024 shots only when the finite-shot anomaly score remains close to a validation-selected security threshold. The quantum scorer is evaluated as an off-path security analytics component rather than part of the haptic critical path. Using a 4,875-record CESNET-TimeSeries24-derived aggregate-flow benchmark, leakage-safe random, entity-group-disjoint, and temporal holdouts, and five trained quantum neural network (QNN) checkpoints per holdout, the primary AS-VQC-95 ($\beta=0.95$) policy averages 129.2, 276.9, and 131.2 shots per record, saving 87.4\%, 73.0\%, and 87.2\% of the uniform 1024-shot baseline (Fixed-1024), respectively. The decision disagreement with analytic (exact-expectation) inference is 0.771\%, 0.409\%, and 0.635\%, lower than both the uniform 128-shot baseline (Fixed-128) and a matched-budget shuffled-allocation control. Fixed-1024 remains more decision-stable, establishing a measurable reliability--resource trade-off rather than cost-free equivalence. A more conservative AS-VQC-99 ($\beta=0.99$) further reduces disagreement while using fewer than 512 average shots across all holdouts. These results show that finite quantum measurements can be treated as an inference resource and concentrated on boundary-sensitive TI-security decisions while exposing checkpoint-dependent escalation under unseen-entity conditions.
\end{abstract}

\section{Introduction}
\label{sec:introduction}

The Tactile Internet (TI) supports interactive control, sensing, and haptic services under stringent requirements for latency, reliability, availability, and security \cite{fettweis2014tactile,promwongsa2021tactile}. Its association with ultra-reliable low-latency communication (URLLC) \cite{popovski2018urllc} and distributed cyber-physical devices makes security evidence both time-sensitive and consequential. Anomaly scores can support monitoring, access control, or mitigation, but aggregate ranking alone does not determine whether individual security decisions remain stable when the underlying inference process is stochastic. This distinction is especially important when a detector is used as one of the evidence sources within a broader zero-trust policy process \cite{nist800207}. We therefore model the quantum anomaly scorer as an off-path analytics component that receives mirrored aggregate-flow telemetry and informs subsequent policy updates. At the same time, deterministic enforcement remains on the TI service path. This split-plane interpretation avoids treating quantum-model execution time as end-to-end haptic or TI latency.
Hybrid quantum--classical classifiers infer measured observables from parameterized quantum circuits \cite{schuld2020circuit}. Analytic simulation returns exact expectation values, whereas finite-shot inference estimates them from a limited measurement budget. Uniformly high shot counts can waste measurements on confident records, while uniformly low counts can destabilize decisions near the operating threshold. We therefore ask whether a trained hybrid QNN can allocate shots adaptively so that easy records terminate early and boundary-sensitive records receive additional measurements.

We evaluate this problem on 4,875 CESNET-TimeSeries24-derived aggregate-flow records \cite{koumar2025cesnet,koumar2024cesnetdataset}. CESNET-TimeSeries24 is not a dedicated TI or haptic-traffic dataset; rather, its operational network telemetry provides a proxy benchmark for evaluating an anomaly-scoring component under record-wise, unseen-entity, and temporal generalization conditions. All fitted preprocessing and pseudo-label statistics are training-only, checkpoint and threshold selection use validation data only, and untouched test records are evaluated under random, entity-group-disjoint, and temporal holdouts. Each selected 12-qubit QNN checkpoint is then reused, without retraining, for both fixed and adaptive-shot inference.
We organize the study around four complementary research questions (RQs) for resource-aware quantum machine learning (QML)-based anomaly detection in TI-oriented security systems. First, we ask whether validation-calibrated adaptive inference can reduce the demand for quantum measurements while improving record-level decision stability relative to low-shot inference (\textit{RQ1}). Second, we test whether uncertainty-targeted allocation outperforms random allocation of the same realized measurement budget (\textit{RQ2}). Third, we examine whether the reliability--measurement trade-off persists across random, entity-group-disjoint, and temporal holdouts (\textit{RQ3}). Fourth, we study how calibration level and trained-checkpoint behavior affect measurement demand, including cases in which many records remain close to the security boundary (\textit{RQ4}).

We make four contributions. First, we formulate finite-shot hybrid quantum anomaly inference as a TI-oriented reliability--resource allocation problem rather than treating shot count as a fixed simulator constant. Second, we introduce AS-VQC, a leakage-safe cumulative policy that calibrates shot-uncertainty margins exclusively on validation data and escalates uncertain predictions through 128, 256, 512, and 1024 shots. Third, we evaluate four fixed-shot baselines, three adaptive calibration levels, and a matched-budget shuffled control across three complementary holdout protocols and five trained QNN checkpoints per holdout. Fourth, AS-VQC-95 saves 72.96--87.39\% of the Fixed-1024 measurement budget while reducing decision disagreement relative to Fixed-128 and the matched-budget shuffled control. AS-VQC-99 further yields lower mean disagreement than Fixed-512 while using fewer than 512 average shots in every holdout. These results characterize a controllable reliability--measurement frontier without claiming quantum computational advantage, Fixed-1024 equivalence, dedicated TI-traffic validation, verified cyberattack detection, or physical-hardware robustness.

\section{Related Work and Positioning}
\label{sec:related_work}

Finite-shot uncertainty and measurement cost are established concerns in variational quantum learning. K\"ubler et al. adapt shot counts during stochastic optimization, Kreplin and Roth study finite-sampling noise in QNNs, Recio-Armengol et al. analyze single-shot QML, and Kim et al. dynamically allocate shots during variational optimization \cite{kubler2020adaptive,kreplin2024finite,recioarmengol2025singleshot,kim2025dynamic}. AS-VQC differs by allocating cumulative measurements after training, per inference record, according to distance from an independently validation-selected decision threshold.

Quantum intrusion-detection studies have explored variational QNNs and QML-based IDS models \cite{gong2022network,abreu2024qmlids}, but primarily emphasize predictive performance rather than measurement allocation. Following guidance on leakage-safe security evaluation and calibration \cite{arp2022dosdonts,guo2017calibration}, our study uses train-only preprocessing, validation-only checkpoint/threshold selection, entity-group and temporal holdouts, and seed-level uncertainty analysis. The matched-budget shuffled control further isolates targeted allocation from simply spending more measurements, which is the specific contribution evaluated here. The present stopping rule specifically assumes a scalar probabilistic output with a deployable decision threshold; we therefore do not claim that AS-VQC directly transfers to quantum algorithms lacking threshold-based outputs.

\section{Tactile-Internet-Oriented Adaptive-Shot Methodology}
\label{sec:methodology}

\begin{figure}[ht]
    \centering
    \includegraphics[width=0.85\linewidth]{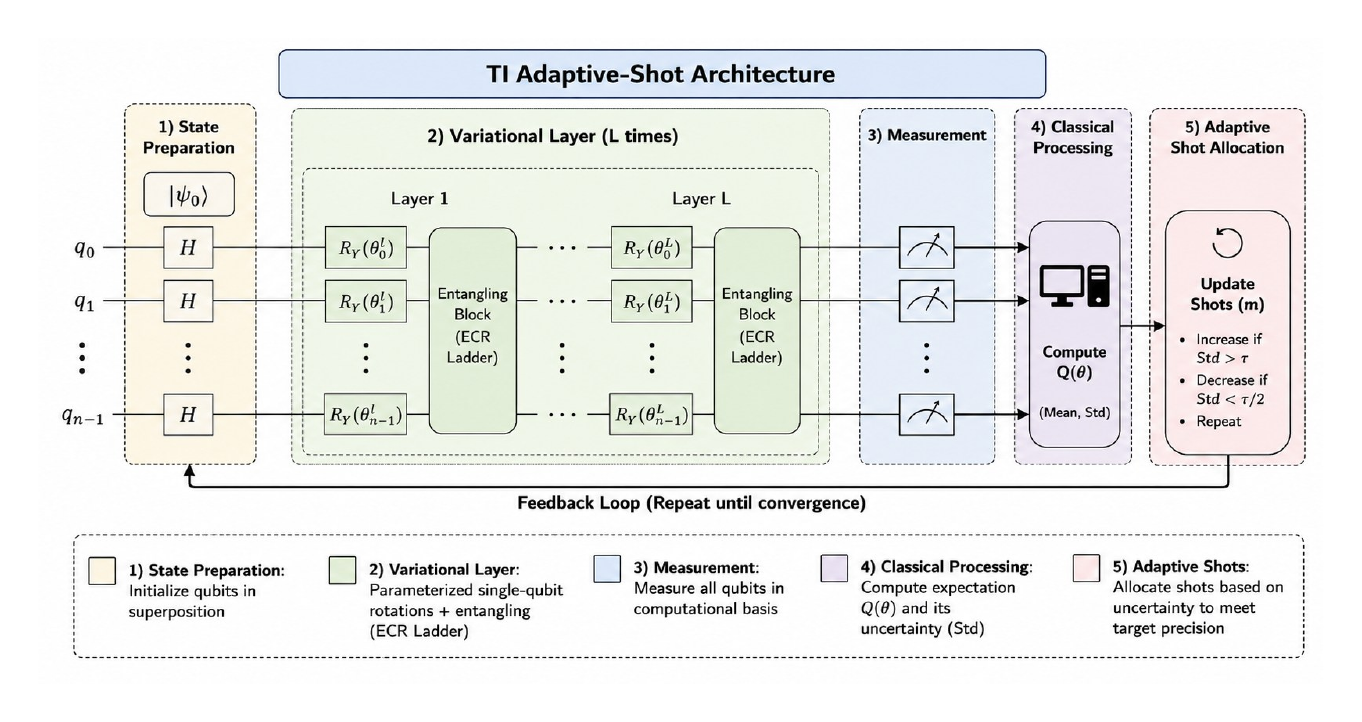}
    \caption{TI-oriented AS-VQC system model with off-path security analytics, validation-calibrated shot escalation, and on-path deterministic enforcement.}
    \label{fig:system_model}
\end{figure}

Figure~\ref{fig:system_model} summarizes the TI-oriented adaptive-shot pipeline. Interactive TI requests remain on the deterministic enforcement path, while mirrored aggregate-flow telemetry is processed off-path by leakage-safe preprocessing, a $12\!\rightarrow\!64\!\rightarrow\!12$ classical embedder, a 12-qubit variational quantum circuit (VQC), and a $2\!\rightarrow\!32\!\rightarrow\!2$ prediction head. Validation area under the receiver operating characteristic curve (ROC-AUC; AUC hereafter) selects the analytic QNN checkpoint, and validation Youden's $J$ determines the fixed threshold $\tau^*$. Finite-shot inference starts at 128 shots and cumulatively escalates to 256, 512, and at most 1024 shots when the prediction remains within the calibrated uncertainty margin. The resulting evidence can inform subsequent policy updates without placing the QNN on the haptic critical path or interpreting shot count as end-to-end TI latency.

\subsection{Benchmark Construction and Leakage-Safe Labels}
\label{sec:method_benchmark}

We construct a 4,875-record benchmark from the 10-minute Internet Protocol (IP) address aggregates of CESNET-TimeSeries24 \cite{koumar2025cesnet}. Records are sampled uniformly without replacement using seed 42 before scaling, filtering, or label construction. Each contains 12 payload-independent flow-level features covering traffic volume, destination diversity, Transmission Control Protocol (TCP)/User Datagram Protocol (UDP) ratios, and directional packet/byte ratios, average duration, and time-to-live. We retain the source time-bin identifier and an entity key for temporal and group-aware partitioning. CESNET-TimeSeries24 is used as an operational proxy benchmark for the candidate TI-security anomaly scorer rather than as a dedicated TI or haptic-traffic capture.

All preprocessing and pseudo-label statistics are fitted on the training partition only. Robust scaling, with zero interquartile range (IQR) replaced by one, is
\begin{equation}
    \widetilde{x}_{ij}=\frac{x_{ij}-m_j}{r_j},
    \label{eq:robust_scale}
\end{equation}
where $x_{ij}$ is feature $j$ of record $i$, $m_j$ and $r_j$ are the training median and IQR, and $\widetilde{x}_{ij}$ is the robust-scaled value.
The training-derived retention interval is
\begin{equation}
    \left[Q_{1j}-1.5\,\mathrm{IQR}_j,\;
          Q_{3j}+1.5\,\mathrm{IQR}_j\right],
    \label{eq:train_outlier_rule}
\end{equation}
where $Q_{1j}$ and $Q_{3j}$ are the training quartiles and $\mathrm{IQR}_j=Q_{3j}-Q_{1j}$ defines the frozen feature-$j$ retention interval, which is applied unchanged to validation and test data. The anomaly score is
\begin{equation}
    a_i=\left\lVert\widetilde{\mathbf{x}}_i\right\rVert_2,
    \label{eq:pseudo_score}
\end{equation}
where $a_i$ is the statistical anomaly score and $\widetilde{\mathbf{x}}_i$ is the robust-scaled feature vector for record $i$.
The 0.85 and 0.95 training-score quantiles define suspicious and high-anomaly states for auditing; the binary task uses
\begin{equation}
    y_i=\mathbb{I}\!\left[a_i\geq
    Q_{0.85}\!\left(\{a_r:r\in\mathcal{T}\}\right)\right],
    \label{eq:binary_pseudolabel}
\end{equation}
where $y_i$ is the binary pseudo-label, $\mathbb{I}[\cdot]$ is the indicator function, $\mathcal{T}$ is the training set, and $Q_{0.85}$ is its 0.85 score quantile.
Thus, validation/test labels depend only on training-derived parameters and represent statistical anomalies rather than verified attacks. Every retained partition must be nonempty and contain both classes.

\subsection{Holdout Protocols}
\label{sec:method_holdouts}

We construct nominal 60/20/20 train/validation/test partitions before applying the training-derived filter. \textit{Random} uses a seeded record permutation; \textit{Entity-group} keeps each entity key in exactly one partition; and \textit{Temporal} assigns ordered time bins chronologically, satisfying
\begin{equation}
\max t_{\mathrm{train}}<\min t_{\mathrm{val}}\leq
\max t_{\mathrm{val}}<\min t_{\mathrm{test}},
\end{equation}
where $t_{\mathrm{train}}$, $t_{\mathrm{val}}$, and $t_{\mathrm{test}}$ denote the retained training, validation, and test time bins, respectively. Random and Group use seeds $\{42,43,44,45,46\}$ for split construction and training randomness; Temporal fixes the chronological split and varies only model/training randomness.

\subsection{Hybrid QNN Architecture}
\label{sec:method_qnn}

The candidate TI-security anomaly scorer contains a classical embedder, a 12-qubit VQC, and a classical prediction head. The embedder maps $12\!\rightarrow\!64\!\rightarrow\!12$ with the Gaussian Error Linear Unit (GELU) activation and produces bounded encoding angles
\begin{equation}
    \mathbf{z}_i=\pi\tanh\!\left(f_{\mathrm{emb}}
    (\widetilde{\mathbf{x}}_i)\right),
    \label{eq:encoding_angles}
\end{equation}
where $f_{\mathrm{emb}}$ is the classical embedder, $\widetilde{\mathbf{x}}_i$ is the scaled input, and $\mathbf{z}_i$ contains the bounded quantum encoding angles.
Starting from $\lvert0\rangle^{\otimes12}$, the $Y$-axis rotation gate $R_Y(z_{ik})$ encodes feature $k$. Each of two variational layers applies a trainable three-angle rotation $\operatorname{Rot}(\phi,\theta,\omega)$, where $\phi$, $\theta$, and $\omega$ are trainable rotation angles, to every qubit, followed by the directed controlled-NOT (CNOT) chain $0\!\rightarrow\!1,\ldots,10\!\rightarrow\!11$. Pauli-$Z$ readout on qubits 0 and 1 gives
\begin{equation}
    \mathbf{m}_i=
    \left(\langle Z_0\rangle_i,\langle Z_1\rangle_i\right),
    \qquad
    p_i=\operatorname{softmax}\!\left(f_{\mathrm{head}}
    (\mathbf{m}_i)\right)_1,
    \label{eq:qnn_readout}
\end{equation}
where $\mathbf{m}_i$ contains the two Pauli-$Z$ expectation values, $f_{\mathrm{head}}$ is the $2\!\rightarrow\!32\!\rightarrow\!2$ GELU head, and $p_i$ is the analytic anomaly probability. The embedder, VQC, and head contain 1,612, 72, and 162 trainable parameters, respectively, for a total of 1,846. The circuit contains 12 encoding gates, 24 trainable rotation gates, and 22 CNOTs. Analytic training uses exact expectation-value simulation on the central processing unit (CPU). For the two-layer directed CNOT chain and $Z_0/Z_1$ readout, backward causal analysis restricts the measured outputs to encoded wires 0--2, enabling exact three-wire reconstruction of the joint readout distribution for the finite-shot experiments below.

\subsection{Training, Validation, and Operating Point}
\label{sec:method_training}

All trainable components are optimized for 20 epochs using Adam with a learning rate of $2\times10^{-3}$, weight decay of $10^{-4}$, batch size 32, gradient clipping at 1.0, and inverse-frequency class weights normalized to unit mean. The checkpoint with the highest validation ROC-AUC is retained, with the first retained in the case of an exact tie; the test set is never used for optimization or model selection. The operating threshold is selected solely from analytic validation probabilities by maximizing Youden's statistic, defined from the true-positive rate (TPR) and false-positive rate (FPR),
\begin{equation}
    \tau^*=\arg\max_{\tau}
    \left\{\operatorname{TPR}_{\mathrm{val}}(\tau)
    -\operatorname{FPR}_{\mathrm{val}}(\tau)\right\},
    \label{eq:youden}
\end{equation}
where $\tau$ is a candidate validation threshold and $\tau^*$ maximizes validation TPR minus FPR; $\tau^*$ is then fixed for all finite-shot test predictions. This separates model and threshold selection from untouched test evaluation.

\subsection{Validation-Calibrated Adaptive-Shot Inference}
\label{sec:method_adaptive}

Adaptive-shot experiments reuse each selected analytically trained QNN checkpoint without retraining. For validation record $i$ and stage $S\in\{128,256,512\}$, the finite-shot error is
\begin{equation}
    e^{(r)}_{i,S}=
    \left|\widehat{p}^{(r)}_{i,S}-p_i\right|,
    \label{eq:validation_shot_error}
\end{equation}
where $e^{(r)}_{i,S}$ is the absolute sampling error for validation record $i$, replicate $r$, and shot count $S$, relative to analytic probability $p_i$.
The stage-specific uncertainty margin is
\begin{equation}
    \delta_{S,\beta}=
    Q_{\beta}\!\left(
    \left\{e^{(r)}_{i,S}\right\}_{i,r}\right),
    \label{eq:adaptive_margin}
\end{equation}
where $\delta_{S,\beta}$ is the stage-$S$ uncertainty margin and $Q_{\beta}$ is the empirical $\beta$-quantile of validation finite-shot errors.
We denote $\beta\in\{0.90,0.95,0.99\}$ by AS-VQC-90, AS-VQC-95, and AS-VQC-99, respectively, so the suffix gives the validation-error calibration percentile; AS-VQC-95 is the primary policy. These margins are empirical validation calibrations and are not presented as formal finite-sample coverage guarantees.
The test-time threshold distance is
\begin{equation}
    d_{i,S}=
    \left|\widehat{p}_{i,S}-\tau^*\right|,
    \label{eq:threshold_distance}
\end{equation}
where $d_{i,S}$ is the finite-shot distance of test record $i$ from the fixed operating threshold $\tau^*$ after $S$ cumulative shots.
The policy stops at stage $S$ when
\begin{equation}
    d_{i,S}>\delta_{S,\beta},
    \label{eq:adaptive_stop}
\end{equation}
where the record stops at stage $S$ when its threshold distance exceeds the validation-calibrated margin for $\beta$; otherwise it acquires more shots. Acquisition is cumulative,
\begin{equation}
    128\rightarrow256\rightarrow512\rightarrow1024,
    \label{eq:cumulative_shots}
\end{equation}
where these are cumulative budgets, so unresolved records receive only the additional shots needed for the next stage; all remaining records stop at 1024.

\subsection{Fixed-Shot and Matched-Budget Controls}
\label{sec:method_controls}

Fixed-128, Fixed-256, Fixed-512, and Fixed-1024 provide uniform-budget baselines. AS-VQC-90 and AS-VQC-99 characterize sensitivity to the validation calibration level around the primary AS-VQC-95 policy. To determine whether adaptive targeting matters beyond average expenditure, Budget-Shuffled-AS95 preserves AS-VQC-95's realized final-shot distribution and average measurement budget but randomly reallocates those budgets across test records. It therefore controls for measurement quantity while removing the uncertainty-targeted assignment rule.
For each test sample, the implementation reconstructs the exact joint distribution $\boldsymbol{\pi}_i=(\pi_{00},\pi_{01},\pi_{10},\pi_{11})$ of measured qubits 0 and 1 using the three-wire causal support. The reconstruction is accepted only when its analytic class probabilities agree with the saved full-model probabilities within a maximum absolute tolerance of $10^{-4}$; the completed matrix achieves a maximum discrepancy of $6.56\times10^{-7}$. Joint multinomial counts are sampled and converted to estimates of $\langle Z_0\rangle$ and $\langle Z_1\rangle$, after which the unchanged trained head produces the finite-shot class probability. The experiment isolates ideal finite-measurement sampling; physical-device noise and execution effects are outside its scope.

\subsection{Metrics and Statistical Aggregation}
\label{sec:method_metrics}

We report AUC and average precision (AP) for ranking; TPR and FPR at $\tau^*$ for operating behavior; Brier score and 15-bin expected calibration error (ECE) for probability quality; and probability mean absolute error (MAE), root mean squared error (RMSE), and bias relative to analytic inference. Record-level decision disagreement is
\begin{equation}
    D=\frac{1}{N}\sum_i
    \mathbb{I}\!\left[
    \mathbb{I}(\widehat{p}_{i}\geq\tau^*)
    \neq\mathbb{I}(p_i\geq\tau^*)\right],
    \label{eq:decision_disagreement}
\end{equation}
where $D$ is the disagreement rate over $N$ test records, comparing each final finite-shot decision $\widehat{p}_i$ with its analytic reference decision $p_i$ at $\tau^*$.
Average measurement demand is
\begin{equation}
    \overline{S}=\frac{1}{N}\sum_i S_i,
    \label{eq:average_shots}
\end{equation}
where $S_i$ is record $i$'s final cumulative shot budget, $N$ is the test-set size, and $\overline{S}$ is the mean measurement demand. Savings relative to Fixed-1024 are
\begin{equation}
    R_S=1-\frac{\overline{S}}{1024},
    \label{eq:shot_savings}
\end{equation}
where $R_S$ is the fractional measurement saving of the adaptive policy relative to the Fixed-1024 reference budget.
For TI-oriented security, threshold-dependent stability matters because false positives can trigger unnecessary restrictions while false negatives can leave anomalous behavior unaddressed. Each holdout contains five trained QNN checkpoints and ten finite-shot realizations per non-analytic policy. Realizations are averaged within each checkpoint, followed by mean $\pm$ sample standard deviation (SD) across five seeds. Paired comparisons use 10,000 bootstrap resamples of the five seed-level differences; measurement repetitions are not treated as independent trained models.

\section{Results}
\label{sec:results}

The completed analysis contains 15 analytically trained QNN checkpoints and 1,200 finite-shot policy realizations, giving 1,215 realization-level rows across the full matrix. The eight finite-shot policies comprise four fixed-shot baselines, AS-VQC-90/95/99, and Budget-Shuffled-AS95. All expected conditions are present, and the reduced-circuit reconstruction validator passes for every checkpoint. The results are interpreted as evidence about measurement allocation and decision reliability for a TI-oriented anomaly-scoring component rather than as end-to-end TI network measurements.

\subsection{Measurement Efficiency and Decision Stability}
\label{sec:results_primary}

Table~\ref{tab:as95_main} reports untouched-test results for the primary AS-VQC-95 policy. Random and Group vary both split construction and training randomness, whereas Temporal fixes the partition and varies only model/training randomness. AS-VQC-95 saves 72.96--87.39\% of the Fixed-1024 budget while keeping disagreement below 0.8\% across all holdouts. Mean AUC remains within $6.4\times10^{-4}$ of analytic inference, although AS-VQC-95 is not treated as equivalent to the analytic or Fixed-1024 reference.

\begin{table}[ht]
\caption{AS-VQC-95 test results (mean $\pm$ sample SD across five checkpoints after within-checkpoint measurement averaging).}
\label{tab:as95_main}
\centering
\scriptsize
\setlength{\tabcolsep}{2.0pt}
\resizebox{\linewidth}{!}{%
\begin{tabular}{@{}lccccccc@{}}
\toprule
\textbf{Holdout} & \textbf{AUC} & \textbf{AP} & \textbf{FPR} &
\textbf{ECE} & \textbf{Disagree} & \textbf{Average Shots} &
\textbf{Saving}\\
\midrule
Random &
$0.9956\pm0.0014$ &
$0.9760\pm0.0050$ &
$0.0299\pm0.0042$ &
$0.0257\pm0.0062$ &
$0.0077\pm0.0012$ &
$129.2\pm0.4$ &
$87.39\pm0.04$\%\\

Group &
$0.9968\pm0.0017$ &
$0.9835\pm0.0092$ &
$0.0346\pm0.0312$ &
$0.0174\pm0.0038$ &
$0.0041\pm0.0010$ &
$276.9\pm327.3$ &
$72.96\pm31.97$\%\\

Temporal &
$0.9980\pm0.0010$ &
$0.9895\pm0.0060$ &
$0.0252\pm0.0104$ &
$0.0211\pm0.0076$ &
$0.0063\pm0.0010$ &
$131.2\pm2.0$ &
$87.19\pm0.20$\%\\
\bottomrule
\end{tabular}%
}
\end{table}

\begin{figure}[ht]
    \centering
    \includegraphics[width=0.75\linewidth]
    {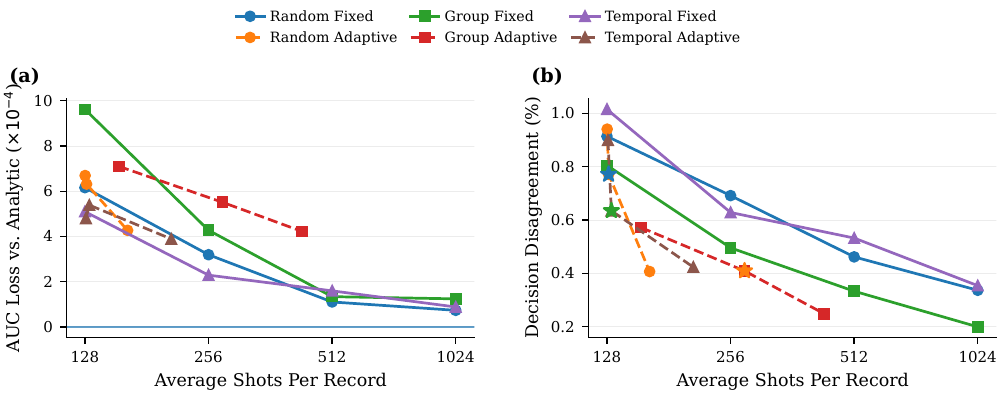}
    \caption{Finite-shot ranking and reliability. (a) ROC-AUC loss relative to analytic inference. (b) Decision disagreement versus average measurement budget; AS-VQC-95 is the primary adaptive policy.}
    \label{fig:ranking_reliability}
\end{figure}

Figure~\ref{fig:ranking_reliability} summarizes ranking preservation and decision reliability within one compact two-panel view. Panel~(a) reports AUC loss relative to analytic inference as measurement demand changes, while panel~(b) places the fixed and adaptive policies on the measurement--reliability plane. Relative to Fixed-128, AS-VQC-95 reduces decision disagreement by 0.142, 0.396, and 0.380 percentage points for Random, Group, and Temporal holdouts. The paired AS-VQC-95-minus-Fixed-128 95\% intervals are $[-0.211,-0.074]$, $[-0.840,-0.056]$, and $[-0.565,-0.129]$ percentage points, respectively, all excluding zero. The corresponding AUC differences are small, and their intervals include zero for all three holdouts. The primary benefit over Fixed-128, therefore, appears at the thresholded decision level rather than as an identifiable ranking improvement.

\subsection{Targeted Allocation Versus Matched-Budget Shuffling}
\label{sec:results_targeting}

\begin{figure}[ht]
    \centering
    \includegraphics[width=0.75\linewidth]
    {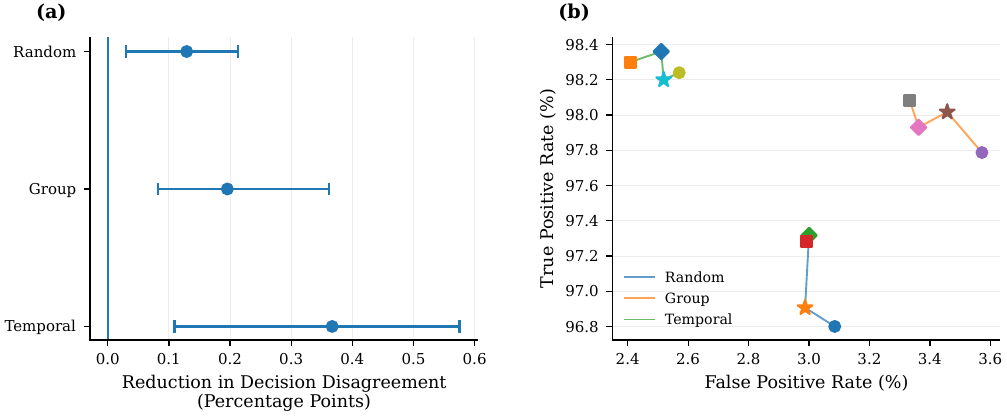}
    \caption{TI-oriented security behavior. (a) AS-VQC-95 disagreement reduction versus matched-budget shuffling, with paired-bootstrap 95\% intervals. (b) TPR/FPR operating points at the fixed validation-selected threshold.}
    \label{fig:targeting_operating}
\end{figure}

The Budget-Shuffled-AS95 control preserves the same realized average shot budget as AS-VQC-95 but removes uncertainty-targeted allocation. AS-VQC-95 lowers decision disagreement by 0.129, 0.196, and 0.367 percentage points relative to the shuffled control for Random, Group, and Temporal holdouts, corresponding to relative reductions of approximately 14.3\%, 32.3\%, and 36.6\%. The paired AS-VQC-95-minus-shuffled disagreement intervals are $[-0.213,-0.030]$, $[-0.362,-0.083]$, and $[-0.575,-0.109]$ percentage points, respectively. Figure~\ref{fig:targeting_operating}(a) therefore shows that where measurements are allocated matters in addition to how many are used. Ranking differences are much smaller: paired AUC differences are $-1.31\times10^{-4}$, $-0.99\times10^{-4}$, and $+0.58\times10^{-4}$, respectively. Thus, the clearest gain from targeted allocation occurs in record-level decision stability rather than aggregate ranking. Figure~\ref{fig:targeting_operating}(b) complements this comparison with the corresponding TI-oriented TPR/FPR operating points for representative fixed-shot and adaptive-shot policies.

\subsection{Adaptive Allocation, Boundary Proximity, and Variability}
\label{sec:results_frontier}
\label{sec:results_boundary}

\begin{figure}[ht]
    \centering
    \includegraphics[width=0.75\linewidth]
    {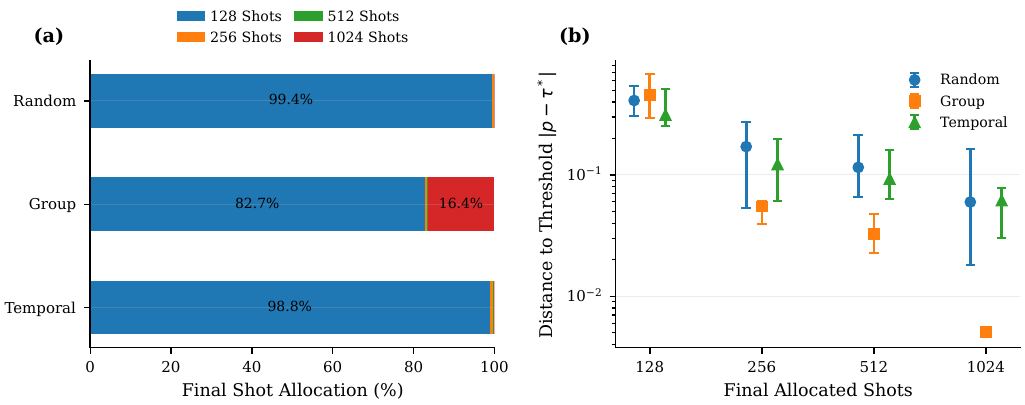}
    \caption{AS-VQC-95 allocation behavior. (a) Fraction of records terminating at each cumulative shot budget. (b) Median analytic threshold distance by final allocation; error bars show pooled interquartile ranges.}
    \label{fig:allocation_boundary}
\end{figure}

AS-VQC-95 allocates most of the measurement expenditure to the initial stage. Figure~\ref{fig:allocation_boundary}(a) shows that 99.44\%, 82.75\%, and 98.83\% of records terminate at 128 shots on average for Random, Group, and Temporal holdouts. The more conservative AS-VQC-99 policy averages 162.6, 432.9, and 207.7 shots with decision disagreement of 0.407\%, 0.248\%, and 0.423\%, respectively. Fixed-512 uses 512 shots per record and produces disagreement of 0.462\%, 0.333\%, and 0.532\%. In observed mean performance, AS-VQC-99 therefore uses fewer measurements and yields lower disagreement than Fixed-512 in every holdout, giving a descriptive Pareto improvement in the measurement-budget/decision-stability plane. AS-VQC-95 remains the pre-specified primary policy.

The escalation mechanism behaves as intended. Figure~\ref{fig:allocation_boundary}(b) groups AS-VQC-95 records by their final shot budget and reports the analytic distance to the fixed validation-selected threshold. For records terminating at 128 versus 1024 shots, the pooled median threshold distance falls from 0.411 to 0.060 for Random, 0.458 to 0.005 for Group, and 0.312 to 0.062 for Temporal. High-shot allocations, therefore, concentrate on records near the security boundary rather than being uniformly distributed across the test set.
The large Group shot-budget SD in Table~\ref{tab:as95_main} is driven by one genuine checkpoint rather than an aggregation error. Group seeds 42--45 average 137.4, 128.1, 128.4, and 128.1 shots per record, whereas seed 46 averages 862.4 shots and sends approximately 81.8\% of records to 1024 shots. Its validation-selected threshold is $\tau^*\approx0.00511$, placing a much larger fraction of test predictions near the operating boundary. We retain this run because it demonstrates that adaptive measurement demand can rise sharply when the learned score distribution and deployed threshold create a dense boundary region. Appendix Fig.~\ref{fig:calibration_resource}(b) visualizes this checkpoint-level variability.

\subsection{Residual Gap to Fixed-1024}
\label{sec:results_fixed1024}

AS-VQC-95 should not be interpreted as equivalent to Fixed-1024. Its paired AUC differences relative to Fixed-1024 are $-0.000558$ (95\% confidence interval (CI) $[-0.000739,-0.000378]$), $-0.000427$ $[-0.000751,-0.000153]$, and $-0.000450$ $[-0.000596,-0.000305]$ for Random, Group, and Temporal, respectively. Decision disagreement is also higher by 0.435, 0.210, and 0.281 percentage points, with all corresponding paired intervals above zero. Fixed-1024, therefore, remains the more stable finite-shot reference. The appropriate conclusion is a controlled reliability--resource trade-off: AS-VQC-95 removes approximately 73--87\% of the 1024-shot budget while accepting a small but measurable finite-shot reliability cost.

\section{Discussion}
\label{sec:discussion}

\textbf{Adaptive Allocation:}\label{sec:discussion_adaptive} The primary result is that uniform measurement expenditure is unnecessary for most evaluated records. Under Random and Temporal holdouts, AS-VQC-95 spends only one to three shots per record above the 128-shot starting budget on average, yet its decision disagreement is significantly lower than Fixed-128. The matched-budget shuffled control further isolates the allocation mechanism: randomly redistributing the same realized shot budget yields greater disagreement across all three holdouts. Thus, the benefit does not arise simply because AS-VQC-95 uses more measurements; it arises because additional measurements are concentrated on predictions that remain close to the operating boundary. For a TI-oriented security plane, this supports treating finite quantum measurements as an inference resource whose allocation should be tied to decision uncertainty rather than applied uniformly.

\textbf{Holdout Robustness:}\label{sec:discussion_holdouts} The adaptive strategy remains useful across record-wise, unseen-entity, and chronological evaluation, but its resource demand is not uniform. Random and Temporal AS-VQC-95 runs remain near the 128-shot floor, whereas Group seed 46 escalates most records to 1024 shots. This result prevents an overly simple claim that adaptive inference always saves approximately 87\% of measurements. Instead, resource efficiency depends on the relationship among the learned probability distribution, validation-selected threshold, and incoming traffic. Figure~\ref{fig:targeting_operating}(b) shows the corresponding TPR/FPR behavior, while Appendix Fig.~\ref{fig:calibration_resource}(a) reports calibration across measurement budgets. For TI-oriented security, average AUC or average shot count should therefore be accompanied by threshold-dependent reliability and the distribution of resource demand.

\textbf{Reliability Resource Interpretation:}\label{sec:discussion_ti} Fixed-1024 remains more decision-stable than AS-VQC-95, so the adaptive method should not be described as free compression or statistically equivalent high-shot inference. Instead, AS-VQC exposes a tunable frontier. AS-VQC-95 prioritizes aggressive measurement savings, while AS-VQC-99 spends more shots to reduce disagreement and, in the observed means, dominates Fixed-512 on the shot/disagreement plane. This trade-off is relevant to an off-path TI security component because additional evidence generation can be concentrated on ambiguous records without claiming that the QNN itself satisfies haptic-loop latency. Shot counts are measurement budgets, not milliseconds, and the present results do not establish quantum-hardware throughput, energy efficiency, or TI/URLLC latency compliance. More broadly, trustworthy resource-aware QML evaluation should report ranking, calibration, operating-point behavior, record-level decision stability, average measurement demand, and checkpoint variability together rather than using any one metric as a deployment proxy.

\section{Limitations}
\label{sec:limitations}

\textbf{Tactile Internet Scope:}
The experiments evaluate a TI-motivated anomaly scorer rather than a complete TI deployment. CESNET-TimeSeries24 is aggregate operational telemetry, not a dedicated haptic/TI corpus; haptic-control loops, URLLC guarantees, end-to-end TI latency, and service availability are therefore not evaluated.

\textbf{Label Validity:}
Targets are training-derived statistical anomaly pseudo-labels from robustly scaled feature norms and empirical quantiles, not expert-verified attacks; the results establish statistical-anomaly discrimination and finite-shot decision behavior rather than verified cyberattack detection.

\textbf{Adaptive Calibration Scope:}
The empirical validation-error quantiles $\delta_{S,\beta}$ provide a leakage-safe stopping rule but no formal coverage or error guarantee; conformal, sequential, or risk-controlled calibration could provide stronger guarantees.

\textbf{Architecture and Task Generality:}
The study evaluates one hybrid QNN architecture and a binary threshold-based anomaly decision. AS-VQC therefore establishes inference-time measurement allocation for this class of probabilistic thresholded models rather than a universal policy for arbitrary quantum algorithms; alternative QNN architectures, regression objectives, or algorithms without scalar decision thresholds require separate validation.

\textbf{Benchmark and Holdout Scope:}
The study uses one 4,875-record CESNET-TimeSeries24 sample and one feature/aggregation design. Random and Group vary split construction and training randomness, whereas Temporal fixes the partition and varies only training randomness; these protocols do not cover all networks, devices, traffic regimes, or concept drift.

\textbf{Quantum Execution Scope:}
Training and analytic reference inference use exact expectation-value simulation, while finite-shot tests sample ideal measurements from fixed checkpoints. Gate/readout noise, decoherence, connectivity, transpilation, queueing, reset time, and calibration drift are excluded; shot count is a measurement-resource metric, not wall-clock latency. Physical-hardware execution is therefore an important next validation step rather than evidence established here.

\textbf{Statistical Scope:}
Each holdout contains five independently trained checkpoints. Ten measurement realizations reduce within-checkpoint estimation noise but do not increase the number of independent trained models; paired-bootstrap intervals over five seed-level differences are interpreted conservatively.

\textbf{Resource Variability Scope:}
Group seed 46 shows that adaptive savings are not guaranteed: when many predictions lie near $\tau^*$, the policy can approach the maximum budget. Deployment may require global budget caps, deferred decisions, step-up actions, or classical fallback procedures.

Overall, the evidence concerns leakage-safe adaptive measurement allocation around analytically trained hybrid QNN checkpoints for a TI-oriented statistical-anomaly benchmark.

\section{Conclusion}
\label{sec:conclusion}

We evaluated AS-VQC as validation-calibrated finite-shot inference for an off-path TI-oriented hybrid QNN. Across random, entity-group-disjoint, and temporal holdouts, AS-VQC-95 saves 72.96--87.39\% of the Fixed-1024 measurement budget while reducing disagreement relative to Fixed-128 and matched-budget random allocation. AS-VQC-99 provides a more conservative operating point, using fewer than 512 average shots while yielding lower mean disagreement than Fixed-512 across all holdouts.
Fixed-1024 remains more stable, and the high-escalation Group checkpoint shows that savings depend on the operating boundary. The result is therefore a reliability--resource trade-off rather than high-shot equivalence: finite measurements can be concentrated on boundary-sensitive security decisions. Future work should examine representative TI/haptic traffic, verified attack labels, formal risk-controlled stopping, noisy hardware, and device-level timing and energy costs.

{\small
\bibliographystyle{unsrtnat}
\bibliography{references/refs}
}

\appendix

\section{Reproducibility Details}
\label{app:reproducibility}

\textbf{Dataset and Candidate Construction:}
The benchmark uses CESNET-TimeSeries24 10-minute IP-address aggregate telemetry \cite{koumar2025cesnet,koumar2024cesnetdataset}. The active source view contains 1,000 source files and 6,863,540 eligible rows. A two-pass uniform sample without replacement with seed 42 selects 4,875 candidate records before fitted scaling, IQR filtering, pseudo-label construction, or split-specific model training. The 12 retained features are \texttt{n\_flows}, \texttt{n\_packets},
\texttt{n\_bytes}, \texttt{n\_dest\_asn}, \texttt{n\_dest\_ports},
\texttt{n\_dest\_ip}, \texttt{tcp\_udp\_ratio\_packets},
\texttt{tcp\_udp\_ratio\_bytes}, \texttt{dir\_ratio\_packets},
\texttt{dir\_ratio\_bytes}, \texttt{avg\_duration}, and \texttt{avg\_ttl}.
The source time bin and entity key are retained for temporal and group-aware partitioning. CESNET-TimeSeries24 is used as an operational proxy benchmark rather than as a dedicated TI or haptic-traffic dataset.

\textbf{Split Construction and Post-Filter Sizes:}
All fitted preprocessing and pseudo-label statistics are training-only. Table~\ref{tab:app_split_counts} reports the retained train/validation/test sizes used by the completed Paper-3 checkpoints. Random and Group use seeds $\{42,43,44,45,46\}$ for split construction and model randomness, whereas Temporal uses one fixed chronological partition across all five training
seeds.
\begin{table}[H]
\caption{Post-filter train/validation/test sizes for the saved Paper-3 splits.}
\label{tab:app_split_counts}
\centering
\scriptsize
\setlength{\tabcolsep}{4.2pt}
\renewcommand{\arraystretch}{0.94}
\begin{tabular}{@{}lrrrr@{}}
\toprule
\textbf{Protocol} & \textbf{Seed} & \textbf{Train} &
\textbf{Validation} & \textbf{Test}\\
\midrule
Random & 42 & 1794 & 602 & 615\\
Random & 43 & 1773 & 602 & 591\\
Random & 44 & 1783 & 591 & 624\\
Random & 45 & 1746 & 592 & 576\\
Random & 46 & 1801 & 595 & 583\\
\midrule
Group & 42 & 1688 & 621 & 653\\
Group & 43 & 1872 & 592 & 522\\
Group & 44 & 1818 & 590 & 506\\
Group & 45 & 1828 & 636 & 592\\
Group & 46 & 1762 & 469 & 669\\
\midrule
Temporal & fixed & 1755 & 561 & 605\\
\bottomrule
\end{tabular}
\end{table}

\textbf{Hybrid QNN Implementation:}
The full model uses a $12\!\rightarrow\!64\!\rightarrow\!12$ classical embedder with GELU followed by $\pi\tanh(\cdot)$ angle bounding. The VQC has 12 qubits and two variational layers. Each layer applies one trainable three-parameter \texttt{Rot} gate per qubit, followed by the directed nearest-neighbor CNOT chain $0\!\rightarrow\!1,\ldots,10\!\rightarrow\!11$. Pauli-$Z$ expectation values on qubits 0 and 1 form the two-dimensional quantum output. The classical head is $2\!\rightarrow\!32\!\rightarrow\!2$ with GELU. The embedder, registered VQC block, and head contain 1,612, 72, and 162 trainable parameters, respectively, for a total of 1,846 parameters. The implementation uses PyTorch and PennyLane interfaces \cite{pennylane_torch,pennylane_measurements}.

\textbf{Optimization, Checkpoint Selection, and Operating Threshold:}
All trainable models use Adam with a learning rate of $2\times10^{-3}$, weight decay of $10^{-4}$, global gradient-norm clipping at 1.0, mini-batches of 32, and 20 epochs. The binary loss uses inverse-frequency training-class weights normalized to unit mean. After each epoch, validation ROC-AUC is computed, and the checkpoint with the largest validation AUC is retained; the first checkpoint is kept in the case of an exact tie. After restoring the selected checkpoint, the operating threshold is chosen only from validation probabilities by maximizing Youden's $J=\mathrm{TPR}-\mathrm{FPR}$ and is applied unchanged to the untouched test partition.

\textbf{Compute Environment and Runtime:}
All reported experiments were executed locally on a Dell Precision 3460 workstation running 64-bit Windows 10 Enterprise. The workstation contains a 12th Gen Intel Core i7-12700 processor with 12 physical cores and 20 logical processors, 32~GB of system memory, and a 500~GB local system drive,
of which approximately 269.23 GB was free when the environment was audited. An NVIDIA T1000 GPU with 4~GB of VRAM and integrated Intel UHD Graphics 770 was available; however, the reported analytic QNN training, analytic inference, and finite-shot measurement simulations were executed on the CPU and did not rely on GPU acceleration. The complete uncached experimental workflow, from candidate preparation through hierarchical aggregation, ran from 00:08:43 to 01:34:39 in the recorded execution, corresponding to approximately 1 hour, 25 minutes, 56 seconds. A subsequent artifact-reusing run that also generated the eight individual diagnostic figures completed in approximately 2 minutes 16 seconds; this shorter runtime reflects the reuse of previously generated QNN checkpoints and adaptive-shot artifacts and should not be interpreted as the
cost of reproducing the full experiment from scratch. Composite figures are generated separately from saved result artifacts and do not require model retraining or repeated finite-shot evaluation.

\textbf{Software Environment:}
Experiments use Python~3.12.10 in a dedicated virtual environment. The repository pins the principal scientific dependencies in \texttt{requirements.txt}, including NumPy~1.26.4, SciPy~1.11.4, pandas~2.2.2, scikit-learn~1.5.1, Matplotlib~3.9.2, PyYAML~6.0.2, tqdm~4.66.5, PyTorch~2.10.0, PennyLane~0.39.0, PennyLane-Lightning~0.39.0, and Autoray~0.6.11. Although the installed PyTorch distribution supports CUDA, the experiments reported in this paper use CPU execution as stated above.

\textbf{Adaptive-Shot Calibration and Evaluation:}
For each trained checkpoint, finite-shot validation realizations determine $\delta_{S,\beta}$ independently for $S\in\{128,256,512\}$ and $\beta\in\{0.90,0.95,0.99\}$. Test acquisition is cumulative: records start at 128 shots, and only unresolved records receive the additional measurements needed to reach 256, 512, and 1024. The primary AS-VQC-95 policy is evaluated alongside AS-VQC-90, AS-VQC-99, four fixed-shot baselines, and Budget-Shuffled-AS95. Ten independent measurement realizations are generated for each non-analytic policy, holdout, and trained checkpoint. Measurement realizations are averaged within each trained checkpoint before variability is summarized across the five independently trained seeds.

\textbf{Artifact Coverage and Statistical Aggregation:}
The completed matrix contains 15 analytically trained QNN checkpoints and 1,200 finite-shot policy realizations, for a total of 1,215 realization-level rows. The seed-level aggregation contains 135 rows, and the summary matrix contains 27 protocol--policy rows. Measurement replicates are averaged within each trained checkpoint before calculating the mean $\pm$ sample SD across five seeds. Paired comparisons use 10,000 percentile-bootstrap resamples of the five seed-level differences. Measurement repetitions are therefore not treated as independent trained models. The maximum reduced-circuit-versus-saved-full-model analytic probability discrepancy is $6.56\times10^{-7}$, below the required $10^{-4}$ validation tolerance.

\textbf{Code Availability, Data Licensing, and Execution Workflow:}
The complete implementation and reproducibility workflow used for the reported experiments are publicly available at \url{https://github.com/msudipto/AdaptiveShot_HybridQAD_Framework}. The repository provides candidate preparation, leakage-safe split construction, QNN training, adaptive-shot calibration, fixed- and adaptive-shot evaluation, hierarchical aggregation, result summarization, and publication-figure generation.
The CESNET-TimeSeries24 source release is cited directly and is distributed under the CC BY 4.0 license. The raw CESNET-TimeSeries24 dataset is not redistributed in the repository; instead, the repository documents the required data preparation and expected input/output structure. The research code introduced for this work is released under the MIT license.
The complete experimental and figure-generation workflow is executed from the repository root using:
\begin{verbatim}
.\run_adaptive-shot_hybrid-qad.ps1
\end{verbatim}
This wrapper performs candidate preparation, leakage-safe Random, Group, and Temporal split construction, training of the 15 source-QNN checkpoints, fixed- and adaptive-shot evaluation, hierarchical aggregation, generation of the individual diagnostic figures, and construction of the four two-panel manuscript figures. Existing experiment artifacts are reused when supported by the corresponding pipeline stage; otherwise, the required outputs are regenerated from the supplied configuration and source code.

\section{Broader Impacts}
\label{app:broader_impacts}

This work is intended to improve resource-aware evaluation of hybrid QML components in Tactile-Internet-oriented security rather than to advocate automated TI decisions based solely on a quantum model. Adaptive measurement allocation can reduce unnecessary inference expenditure and make record-level uncertainty visible, while the matched-budget shuffled control helps distinguish intelligent targeting from simple increases in measurement quantity. Reporting the full reliability--measurement frontier may support more transparent decisions about whether and where finite-shot quantum inference is operationally justified.
The study nevertheless uses statistical-anomaly pseudo-labels rather than verified attack annotations. If deployed without independent contextual signals, errors near the operating threshold could create unnecessary alerts, restrictions, investigations, or missed anomalies. Because TI applications may couple network decisions to interactive or control-sensitive services, disruptive security actions should not depend on a single anomaly score without appropriate policy safeguards, auditing, and human or administrative review. Flow-level telemetry also raises privacy and governance concerns even when payload content is not inspected. The reported shot savings should not be interpreted as evidence of physical-device energy savings, compliance with haptic-loop latency, or safe autonomous deployment.

\section{Supplementary Adaptive-Shot Diagnostics}
\label{app:supplementary_diagnostics}

\subsection{Calibration and Checkpoint-Level Resource Demand}
\label{app:calibration_resource}

\begin{figure}[H]
    \centering
    \includegraphics[width=0.75\linewidth]
    {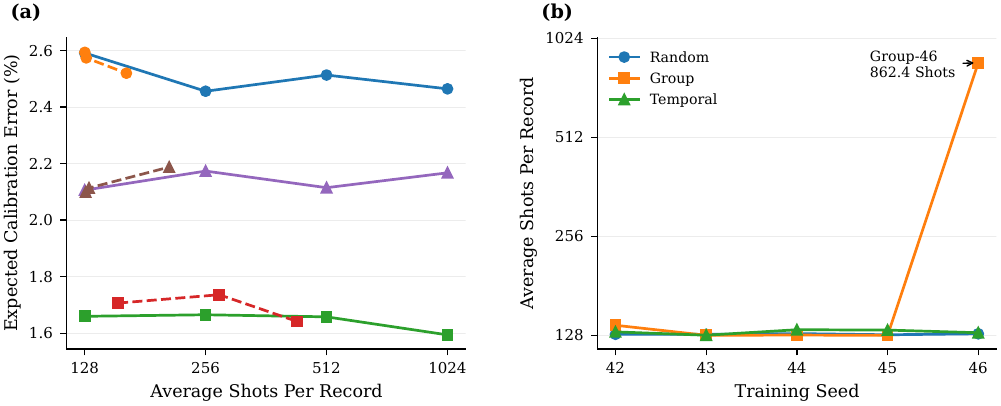}
    \caption{Calibration and checkpoint-level resource behavior. (a) Expected calibration error across fixed-shot and adaptive-shot measurement budgets. (b) AS-VQC-95 average measurement demand for each trained checkpoint, highlighting the high-escalation Group seed-46 run.}
    \label{fig:calibration_resource}
\end{figure}

Figure~\ref{fig:calibration_resource}(a) reports expected calibration error as a function of average measurement budget. Calibration does not collapse to the same ordering as decision disagreement, reinforcing that probability quality, ranking, and threshold-level stability should be evaluated separately. Figure~\ref{fig:calibration_resource}(b) reports AS-VQC-95 average shots for every trained checkpoint. Random and Temporal checkpoints remain close to the 128-shot floor, while Group seed 46 requires substantially more measurement effort. The two panels jointly show that resource-aware inference should be evaluated in terms of both predictive calibration and checkpoint-level measurement demand rather than average shot count alone.

\subsection{Paired Decision-Stability Summary}
\label{app:paired_summary}

Table~\ref{tab:app_paired_disagreement} reports the primary paired differences in decision disagreement. Negative values favor AS-VQC-95 for the Fixed-128 and matched-budget shuffled comparisons; positive values in the Fixed-1024 comparison indicate that the higher-shot baseline remains more stable.

\begin{table}[H]
\caption{Paired AS-VQC-95 decision-disagreement differences with percentile 95\% bootstrap intervals over five seed-level differences. Values are percentage points.}
\label{tab:app_paired_disagreement}
\centering
\scriptsize
\setlength{\tabcolsep}{4.0pt}
\begin{tabular}{@{}llcc@{}}
\toprule
\textbf{Holdout} & \textbf{Comparator} &
\textbf{Mean $\Delta$Disagree} & \textbf{95\% CI}\\
\midrule
Random & Fixed-128 & $-0.142$ & $[-0.211,\ -0.074]$\\
Group & Fixed-128 & $-0.396$ & $[-0.840,\ -0.056]$\\
Temporal & Fixed-128 & $-0.380$ & $[-0.565,\ -0.129]$\\
\midrule
Random & Budget-Shuffled-AS95 & $-0.129$ & $[-0.213,\ -0.030]$\\
Group & Budget-Shuffled-AS95 & $-0.196$ & $[-0.362,\ -0.083]$\\
Temporal & Budget-Shuffled-AS95 & $-0.367$ & $[-0.575,\ -0.109]$\\
\midrule
Random & Fixed-1024 & $+0.435$ & $[0.376,\ 0.497]$\\
Group & Fixed-1024 & $+0.210$ & $[0.107,\ 0.294]$\\
Temporal & Fixed-1024 & $+0.281$ & $[0.255,\ 0.307]$\\
\bottomrule
\end{tabular}
\end{table}

\clearpage
\input{checklist.tex}

\end{document}

%% file: checklist.tex
\section*{NeurIPS Paper Checklist}

\begin{enumerate}

\item {\bf Claims}
    \item[] Question: Do the main claims made in the abstract and introduction accurately reflect the paper's contributions and scope?
    \item[] Answer: \answerYes{}
    \item[] Justification: The Abstract and Section~\ref{sec:introduction} state the paper's scope, contributions, TI-oriented off-path interpretation, and reliability--resource claims consistently with the empirical findings in Section~\ref{sec:results}. The manuscript explicitly avoids claims of quantum advantage, Fixed-1024 equivalence, dedicated TI-traffic validation, verified attack detection, or physical-hardware robustness.

\item {\bf Limitations}
    \item[] Question: Does the paper discuss the limitations of the work performed by the authors?
    \item[] Answer: \answerYes{}
    \item[] Justification: Section~\ref{sec:limitations} explicitly discusses the TI application scope, pseudo-label validity, empirical adaptive calibration, benchmark/holdout scope, ideal finite-shot simulation, statistical scope, and checkpoint-dependent resource variability.

\item {\bf Theory assumptions and proofs}
    \item[] Question: For each theoretical result, does the paper provide the full set of assumptions and a complete (and correct) proof?
    \item[] Answer: \answerNA{}
    \item[] Justification: The paper does not present theorems, lemmas, or proof-based theoretical results. The numbered equations define the model, stopping rule, operating threshold, and evaluation metrics used in the empirical study.

\item {\bf Experimental result reproducibility}
    \item[] Question: Does the paper fully disclose all the information needed to reproduce the main experimental results of the paper to the extent that it affects the main claims and/or conclusions of the paper (regardless of whether the code and data are provided or not)?
    \item[] Answer: \answerYes{}
    \item[] Justification:
    The complete code used for the reported experiments is publicly available at \url{https://github.com/msudipto/AdaptiveShot_HybridQAD_Framework}. The repository provides the software requirements, experiment configuration, candidate-preparation code, leakage-safe split construction, QNN training, adaptive-shot calibration and evaluation, hierarchical aggregation, summary generation, and publication-figure scripts, together with the end-to-end execution command. CESNET-TimeSeries24 is publicly available through the cited source release; the repository documents data preparation and expected paths rather than redistributing the raw dataset.

\item {\bf Open access to data and code}
    \item[] Question: Does the paper provide open access to the data and code, with sufficient instructions to faithfully reproduce the main experimental results, as described in supplemental material?
    \item[] Answer: \answerYes{}
    \item[] Justification: The CESNET-TimeSeries24 source data are publicly available through the cited Zenodo release, and the supplementary code package accompanying the paper provides the configuration, scripts, and commands for candidate preparation, splitting, QNN training, adaptive-shot evaluation, aggregation, and figure generation as described in Appendix~\ref{app:reproducibility}.
    \item[] Guidelines:
    \begin{itemize}
        \item The answer \answerNA{} means that the paper does not include experiments requiring code.
        \item Please see the NeurIPS code and data submission guidelines (\url{https://neurips.cc/public/guides/CodeSubmissionPolicy}) for more details.
        \item While we encourage the release of code and data, we understand that this might not be possible, so \answerNo{} is an acceptable answer. Papers cannot be rejected simply for not including code unless code is central to the contribution (e.g., a new open-source benchmark).
        \item The instructions should contain the exact command and environment needed to run to reproduce the results. See the NeurIPS code and data submission guidelines (\url{https://neurips.cc/public/guides/CodeSubmissionPolicy}) for more details.
        \item The authors should provide instructions on data access and preparation, including how to access the raw data, preprocessed data, intermediate data, and generated data, etc.
        \item The authors should provide scripts to reproduce all experimental results for the new proposed method and baselines. If only a subset of experiments are reproducible, they should state which ones are omitted from the script and why.
        \item At submission time, to preserve anonymity, the authors should release anonymized versions (if applicable).
        \item Providing as much information as possible in supplemental material (appended to the paper) is recommended, but including URLs to data and code is permitted.
    \end{itemize}

\item {\bf Experimental setting/details}
    \item[] Question: Does the paper specify all the training and test details (e.g., data splits, hyperparameters, how they were chosen, type of optimizer) necessary to understand the results?
    \item[] Answer: \answerYes{}
    \item[] Justification:
    Sections~\ref{sec:method_benchmark}--\ref{sec:method_metrics} report the split construction, training-only preprocessing, model architecture, optimizer and hyperparameters, checkpoint rule, validation-selected threshold, finite-shot stages, adaptive calibration levels, controls, metrics, seeds, and statistical aggregation. Appendix~\ref{app:reproducibility} provides the software and compute environment, while the exact configuration and executable implementation are publicly available at \url{https://github.com/msudipto/AdaptiveShot_HybridQAD_Framework}.

\item {\bf Experiment statistical significance}
    \item[] Question: Does the paper report error bars suitably and correctly defined or other appropriate information about the statistical significance of the experiments?
    \item[] Answer: \answerYes{}
    \item[] Justification: Table~\ref{tab:as95_main} reports mean $\pm$ sample SD across five independently trained checkpoints after averaging measurement realizations within each checkpoint, and the main text reports paired percentile 95\% bootstrap intervals from 10,000 resamples of the five seed-level differences. Section~\ref{sec:method_metrics} defines the sources of variability and aggregation order.
    \item[] Guidelines:
    \begin{itemize}
        \item The answer \answerNA{} means that the paper does not include experiments.
        \item The authors should answer \answerYes{} if the results are accompanied by error bars, confidence intervals, or statistical significance tests, at least for the experiments that support the main claims of the paper.
        \item The factors of variability that the error bars are capturing should be clearly stated (for example, train/test split, initialization, random drawing of some parameter, or overall run with given experimental conditions).
        \item The method for calculating the error bars should be explained (closed form formula, call to a library function, bootstrap, etc.)
        \item The assumptions made should be given (e.g., Normally distributed errors).
        \item It should be clear whether the error bar is the standard deviation or the standard error of the mean.
        \item It is OK to report 1-sigma error bars, but one should state it. The authors should preferably report a 2-sigma error bar than state that they have a 96\% CI, if the hypothesis of Normality of errors is not verified.
        \item For asymmetric distributions, the authors should be careful not to show in tables or plots symmetric error bars that would yield results that are out of range (e.g., negative error rates).
        \item If error bars are reported in tables or plots, the authors should explain in the text how they were calculated and reference the corresponding figures or tables in the text.
    \end{itemize}

\item {\bf Experiments compute resources}
    \item[] Question: For each experiment, does the paper provide sufficient information on the computer resources (type of compute workers, memory, time of execution) needed to reproduce the experiments?
    \item[] Answer: \answerYes{}
    \item[] Justification:
    Appendix~\ref{app:reproducibility} reports the experimental compute environment: a Dell Precision 3460 workstation with a 12th Gen Intel Core i7-12700 processor (12 physical cores and 20 logical processors), 31.7~GB of RAM, approximately 474.65~GB of local storage, and an NVIDIA T1000 GPU with 4~GB VRAM. The reported QNN training, analytic inference, and finite-shot simulations were executed on CPU using Python~3.12.10. The complete uncached experimental workflow required approximately 1~hour 25~minutes 56~seconds.
    
\item {\bf Code of ethics}
    \item[] Question: Does the research conducted in the paper conform, in every respect, with the NeurIPS Code of Ethics \url{https://neurips.cc/public/EthicsGuidelines}?
    \item[] Answer: \answerYes{}
    \item[] Justification:
    The study uses publicly released aggregate-flow telemetry, does not involve human subjects, and discusses security, privacy, deployment, and misuse considerations in Appendix~\ref{app:broader_impacts}. No sensitive packet payloads or private participant data are collected or released.

\item {\bf Broader impacts}
    \item[] Question: Does the paper discuss both potential positive societal impacts and negative societal impacts of the work performed?
    \item[] Answer: \answerYes{}
    \item[] Justification: Appendix~\ref{app:broader_impacts} discusses potential benefits of resource-aware security inference as well as risks from false decisions, unnecessary restrictions, privacy/governance concerns, and overinterpreting simulated shot savings as hardware or TI-latency guarantees.
    
\item {\bf Safeguards}
    \item[] Question: Does the paper describe safeguards that have been put in place for responsible release of data or models that have a high risk for misuse (e.g., pre-trained language models, image generators, or scraped datasets)?
    \item[] Answer: \answerNA{}
    \item[] Justification: The work does not release a high-risk generative model, scraped media dataset, or other asset requiring controlled access or safety filtering. The released material consists of research code, configurations, and derived experiment artifacts for the reported anomaly-scoring study.

\item {\bf Licenses for existing assets}
    \item[] Question: Are the creators or original owners of assets (e.g., code, data, models), used in the paper, properly credited and are the license and terms of use explicitly mentioned and properly respected?
    \item[] Answer: \answerYes{}
    \item[] Justification:
    The CESNET-TimeSeries24 creators and source publication are explicitly cited, and the dataset release is distributed under CC BY 4.0. The raw dataset is not redistributed by this work. The public reproducibility repository at \url{https://github.com/msudipto/AdaptiveShot_HybridQAD_Framework} documents the data dependency and third-party software requirements. The new research code introduced for this study is distributed under the MIT license.

\item {\bf New assets}
    \item[] Question: Are new assets introduced in the paper well documented and is the documentation provided alongside the assets?
    \item[] Answer: \answerYes{}
    \item[] Justification:
    The AS-VQC implementation, experiment configuration, candidate-preparation code, leakage-safe splitting, QNN training, fixed- and adaptive-shot evaluation, hierarchical aggregation, result summaries, and figure-generation workflow are publicly released and documented at \url{https://github.com/msudipto/AdaptiveShot_HybridQAD_Framework}. The repository includes software requirements, setup instructions, execution commands, expected artifact structure, and licensing information.

\item {\bf Crowdsourcing and research with human subjects}
    \item[] Question: For crowdsourcing experiments and research with human subjects, does the paper include the full text of instructions given to participants and screenshots, if applicable, as well as details about compensation (if any)? 
    \item[] Answer: \answerNA{}
    \item[] Justification: The research does not involve crowdsourcing, participant recruitment, user studies, or other experiments with human subjects.

\item {\bf Institutional review board (IRB) approvals or equivalent for research with human subjects}
    \item[] Question: Does the paper describe potential risks incurred by study participants, whether such risks were disclosed to the subjects, and whether Institutional Review Board (IRB) approvals (or an equivalent approval/review based on the requirements of your country or institution) were obtained?
    \item[] Answer: \answerNA{}
    \item[] Justification: The research does not involve human subjects or crowdsourcing, so IRB or equivalent human-subjects review is not applicable.

\item {\bf Declaration of LLM usage}
    \item[] Question: Does the paper describe the usage of LLMs if it is an important, original, or non-standard component of the core methods in this research? Note that if the LLM is used only for writing, editing, or formatting purposes and does \emph{not} impact the core methodology, scientific rigor, or originality of the research, declaration is not required.
    \item[] Answer: \answerNA{}
    \item[] Justification:
    Large language models are not an important, original, or non-standard component of the AS-VQC method, QNN training, adaptive-shot inference, finite-shot evaluation, or statistical analysis. Any language-model assistance used for writing, editing, or formatting does not affect the core methodology, scientific results, or originality of the research.

\end{enumerate}

%% file: references/refs.bib
@article{fettweis2014tactile,
  author  = {Fettweis, Gerhard P.},
  title   = {The Tactile Internet: Applications and Challenges},
  journal = {IEEE Vehicular Technology Magazine},
  volume  = {9},
  number  = {1},
  pages   = {64--70},
  year    = {2014},
  doi     = {10.1109/MVT.2013.2295069}
}

@article{promwongsa2021tactile,
  author  = {Promwongsa, Nattakorn and Ebrahimzadeh, Amin and Naboulsi, Diala and
             Kianpisheh, Somayeh and Belqasmi, Fatna and Glitho, Roch and
             Crespi, Noel and Alfandi, Omar},
  title   = {A Comprehensive Survey of the Tactile Internet: State-of-the-Art
             and Research Directions},
  journal = {IEEE Communications Surveys \& Tutorials},
  volume  = {23},
  number  = {1},
  pages   = {472--523},
  year    = {2021},
  doi     = {10.1109/COMST.2020.3025995}
}

@article{popovski2018urllc,
  author  = {Popovski, Petar and Nielsen, Jimmy J. and Stefanovic, Cedomir and
             {de Carvalho}, Elisabeth and Str{\"o}m, Erik and
             Trillingsgaard, Kasper F. and Bana, Alexandru-Sabin and
             Kim, Dong Min and Kotaba, Radoslaw and Park, Jihong and
             S{\o}rensen, Ren{\'e} B.},
  title   = {Wireless Access for {URLLC}: Principles and Building Blocks},
  journal = {IEEE Network},
  volume  = {32},
  number  = {2},
  pages   = {16--23},
  year    = {2018},
  doi     = {10.1109/MNET.2018.1700258}
}

@techreport{nist800207,
  author      = {Rose, Scott and Borchert, Oliver and Mitchell, Stuart and
                 Connelly, Sean},
  title       = {Zero Trust Architecture},
  institution = {National Institute of Standards and Technology},
  number      = {NIST Special Publication 800-207},
  year        = {2020},
  doi         = {10.6028/NIST.SP.800-207}
}

@article{koumar2025cesnet,
  author  = {Koumar, Josef and Hynek, Karel and {\v{C}}ejka, Tom{\'a}{\v{s}} and
             {\v{S}}i{\v{s}}ka, Pavel},
  title   = {{CESNET-TimeSeries24}: Time Series Dataset for Network Traffic
             Anomaly Detection and Forecasting},
  journal = {Scientific Data},
  volume  = {12},
  pages   = {338},
  year    = {2025},
  doi     = {10.1038/s41597-025-04603-x}
}

@misc{koumar2024cesnetdataset,
  author    = {Koumar, Josef and Hynek, Karel and {\v{C}}ejka, Tom{\'a}{\v{s}} and
               {\v{S}}i{\v{s}}ka, Pavel},
  title     = {{CESNET-TimeSeries24}: Time Series Dataset for Network Traffic
               Anomaly Detection and Forecasting},
  publisher = {Zenodo},
  year      = {2024},
  doi       = {10.5281/zenodo.13382427}
}

@article{schuld2020circuit,
  author  = {Schuld, Maria and Bocharov, Alex and Svore, Krysta M. and
             Wiebe, Nathan},
  title   = {Circuit-Centric Quantum Classifiers},
  journal = {Physical Review A},
  volume  = {101},
  number  = {3},
  pages   = {032308},
  year    = {2020},
  doi     = {10.1103/PhysRevA.101.032308}
}

@article{kubler2020adaptive,
  author  = {K{\"u}bler, Jonas M. and Arrasmith, Andrew and Cincio, {\L}ukasz and
             Coles, Patrick J.},
  title   = {An Adaptive Optimizer for Measurement-Frugal Variational Algorithms},
  journal = {Quantum},
  volume  = {4},
  pages   = {263},
  year    = {2020},
  doi     = {10.22331/q-2020-05-11-263}
}

@article{kim2025dynamic,
  author  = {Kim, Youngmin and Jang, Enhyeok and Kim, Hyungseok and
             Choi, Seungwoo and Lee, Changheon and Kim, Donghwi and
             Kyoung, Woomin and Shin, Kyujin and Ro, Won Woo},
  title   = {Distribution-Adaptive Dynamic Shot Optimization for Variational
             Quantum Algorithms},
  journal = {Physical Review Research},
  volume  = {7},
  number  = {4},
  pages   = {043253},
  year    = {2025}
}

@article{kreplin2024finite,
  author  = {Kreplin, David A. and Roth, Marco},
  title   = {Reduction of Finite Sampling Noise in Quantum Neural Networks},
  journal = {Quantum},
  volume  = {8},
  pages   = {1385},
  year    = {2024},
  doi     = {10.22331/q-2024-06-25-1385}
}

@article{recioarmengol2025singleshot,
  author  = {Recio-Armengol, Erik and Eisert, Jens and Meyer, Johannes Jakob},
  title   = {Single-Shot Quantum Machine Learning},
  journal = {Physical Review A},
  volume  = {111},
  number  = {4},
  pages   = {042420},
  year    = {2025},
  doi     = {10.1103/PhysRevA.111.042420}
}

@article{gong2022network,
  author  = {Gong, Changqing and Guan, Weiqi and Gani, Abdullah and Qi, Han},
  title   = {Network Attack Detection Scheme Based on Variational Quantum
             Neural Network},
  journal = {The Journal of Supercomputing},
  volume  = {78},
  number  = {15},
  pages   = {16876--16897},
  year    = {2022},
  doi     = {10.1007/s11227-022-04542-z}
}

@article{abreu2024qmlids,
  author  = {Abreu, Diego and Rothenberg, Christian Esteve and
             Abel{\'e}m, Antonio},
  title   = {{QML-IDS}: Quantum Machine Learning Intrusion Detection System},
  journal = {arXiv preprint arXiv:2410.16308},
  year    = {2024}
}

@inproceedings{arp2022dosdonts,
  author    = {Arp, Daniel and Quiring, Erwin and Pendlebury, Feargus and
               Warnecke, Alexander and Pierazzi, Fabio and
               Wressnegger, Christian and Cavallaro, Lorenzo and
               Rieck, Konrad},
  title     = {Dos and Don'ts of Machine Learning in Computer Security},
  booktitle = {31st USENIX Security Symposium (USENIX Security 22)},
  pages     = {3971--3988},
  year      = {2022},
  address   = {Boston, MA},
  publisher = {USENIX Association}
}

@inproceedings{guo2017calibration,
  author    = {Guo, Chuan and Pleiss, Geoff and Sun, Yu and
               Weinberger, Kilian Q.},
  title     = {On Calibration of Modern Neural Networks},
  booktitle = {Proceedings of the 34th International Conference on Machine Learning},
  series    = {Proceedings of Machine Learning Research},
  volume    = {70},
  pages     = {1321--1330},
  year      = {2017},
  publisher = {PMLR}
}

@misc{pennylane_torch,
  author = {{PennyLane}},
  title  = {PyTorch Interface},
  year   = {2026},
  note   = {PennyLane Documentation}
}

@misc{pennylane_measurements,
  author = {{PennyLane}},
  title  = {Quantum Measurements},
  year   = {2026},
  note   = {PennyLane Documentation}
}
